\documentclass[%
 reprint,
 amsmath,amssymb,
 aps,
]{revtex4-2}
\usepackage{graphicx}
\usepackage{dcolumn}
\usepackage{bm}
\usepackage{braket}
\usepackage{xcolor}
\usepackage{float}
\renewcommand{\vec}[1]{\boldsymbol{\mathbf{#1}}}
\usepackage{ulem} 
\usepackage{hyperref}

\begin{document}

\title{Pauli blocking of light scattering in degenerate fermions}

\author{Yair Margalit}
\affiliation{Research Laboratory of Electronics, MIT-Harvard Center for Ultracold Atoms, and Department of Physics, Massachusetts Institute of Technology, Cambridge, Massachusetts 02139, USA}

\author{Yu-Kun Lu}%
\affiliation{Research Laboratory of Electronics, MIT-Harvard Center for Ultracold Atoms, and Department of Physics, Massachusetts Institute of Technology, Cambridge, Massachusetts 02139, USA}

\author{Furkan \c{C}a\u{g}r\i{} Top}
\affiliation{Research Laboratory of Electronics, MIT-Harvard Center for Ultracold Atoms, and Department of Physics, Massachusetts Institute of Technology, Cambridge, Massachusetts 02139, USA}

\author{Wolfgang Ketterle}
\affiliation{Research Laboratory of Electronics, MIT-Harvard Center for Ultracold Atoms, and Department of Physics, Massachusetts Institute of Technology, Cambridge, Massachusetts 02139, USA}


\date{\today}

\begin{abstract}
\end{abstract}

\maketitle


\textbf{Pauli blocking of spontaneous emission is responsible for the stability of atoms. Higher electronic orbitals cannot decay to lower-lying states if they are already occupied --- this is Pauli blocking due to occupation of internal states. Pauli blocking also occurs when free atoms scatter light elastically (Rayleigh scattering) and the final external momentum states are already occupied. A suppression of the total rate of light scattering requires a quantum-degenerate Fermi gas with a Fermi energy larger than the photon recoil energy. This has been predicted more than 30 years ago, but never realized. Here we report the creation of a dense Fermi gas of ultracold lithium atoms and show that at low temperatures light scattering is suppressed. We also explore the suppression of inelastic light scattering when two colliding atoms emit light shifted in frequency.}

\begin{figure*}
\includegraphics[width=2\columnwidth]{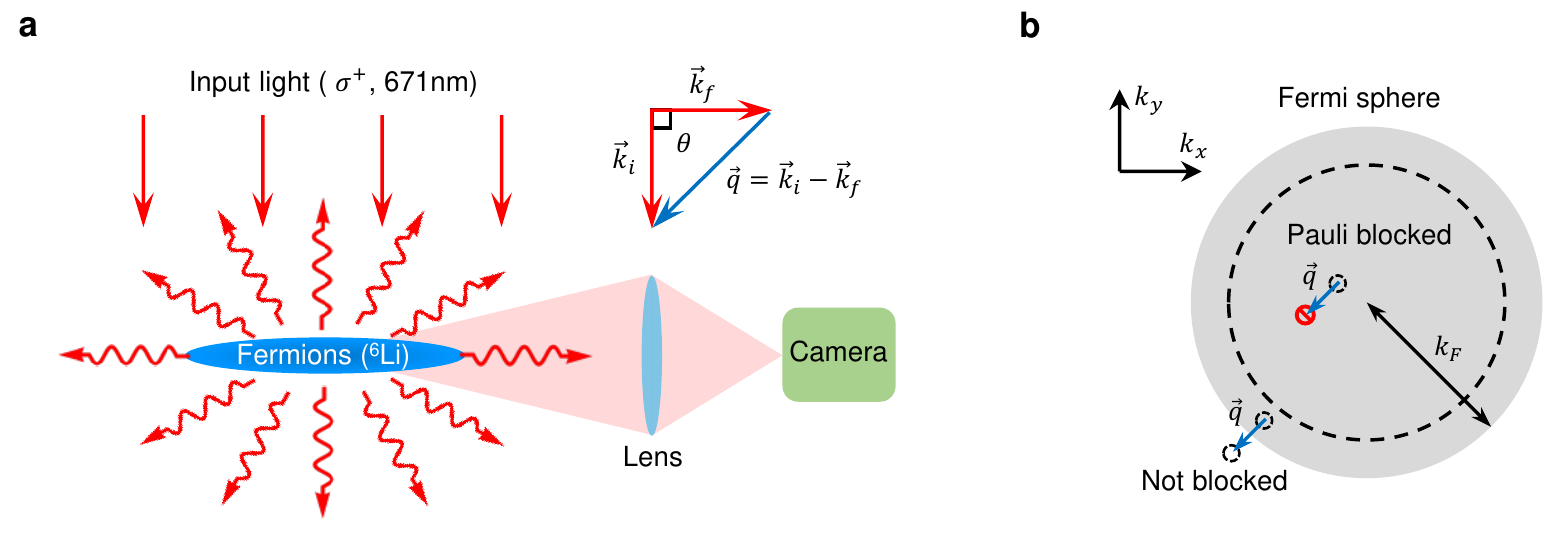}
\caption{\label{fig1} \textbf{Schematic of the experiment}. \textbf{a}, Degenerate fermionic lithium atoms are confined in an optical dipole trap and illuminated with a laser beam. Scattered photons impart momentum transfers of $\hbar q$ to the atoms and are detected at a scattering angle of 90 degrees. \textbf{b}, Mechanism of Pauli blocking in degenerate Fermi gases. At temperature $T=0$ the atoms occupy a Fermi sphere in momentum space with radius $\hbar k_F$. For $q\ll k_F$, atoms can scatter light only from the outer shell of the Fermi sphere (of width $\hbar q$) where they can reach an unoccupied final momentum state. No scattering is possible for atoms within the dashed circle.}
\end{figure*}

Starting in 1990, several papers have predicted the suppression of light scattering in ultracold Fermi gases \cite{HELMERSON:90,PhysRevA.52.3033,Busch_1998,DeMarco1998,PhysRevLett.82.4741,Shuve_2009,PhysRevA.79.033602,PhysRevA.84.043825}. The basic phenomenon is show in Fig. \ref{fig1}. Light scattering between photon states with wavevectors $k_i$ and $k_f$ transfers momentum $\hbar q = \hbar (k_i - k_f)= 2\hbar k \sin\theta/2$, where $\hbar k$ is the photon momentum and $\theta$ the scattering angle. When the Fermi momentum $\hbar k_F$ of a zero temperature Fermi gas is larger than the momentum transfer $\hbar q$, light scattering is strongly suppressed and can occur only near the Fermi surface, whereas for temperatures $T \geq T_F$, the scattering rate per atom approaches the independent atom limit. This smooth transition versus temperature has been theoretically studied, including averaging over the inhomogeneous density distribution of a harmonically trapped atom cloud \cite{Shuve_2009}.

Experiments on ultracold atoms have deepened our understanding of basic physical phenomena by realizing paradigmatic idealized situations where the phenomenon is observed in its most direct and transparent way. These realizations then become building blocks for more complex systems or quantum simulations. Examples include the realization of Bose-Einstein condensation, the BEC-BCS cross over in fermions, band structure phenomena of non-interacting atoms in optical lattices, and Mott insulators in optical lattices \cite{RevModPhys.80.885}. Here we study, in a highly idealized situation, how ultracold fermions scatter light. Recently, we have been able to prepare ultracold fermions at unprecedented densities ($n=3\times 10^{15} cm^{-3}$) \cite{ccaugri2020spin}, where the Fermi energy is 50 times higher than the photon recoil energy $\hbar^2 k_i^2/2m$ of 73.9\,kHz (where m is the lithium atomic mass). Using this sample, we have now performed light scattering experiments in the simplest possible limit, at detunings $\Delta$ from the atomic resonance by up to 500 GHz, or 85,000 linewidths $\Gamma$. Therefore, despite high atomic densities $n \approx 3.4/ \lambdabar^3$ and high resonant optical densities $ 6 \pi n \lambdabar ^2 l \approx 67,000$ (where $\lambdabar = 1/k$, and $l$ is the length of the atom cloud), we realize the limit where both the absorptive and dispersive parts of the index of refraction are negligible. In general, optical properties become complicated in the regime of high densities due to strong Lorentz-Lorentz corrections \cite{jackson1999classical} and dipolar interactions between the atoms \cite{PhysRevA.94.023612}. These corrections are often expanded in the parameter $n \alpha$, where $\alpha$ is the atomic polarizability, given for a two-level atom by $\alpha = 6\pi \lambdabar^3 \Gamma/ (\Delta + i\Gamma)$. At our detunings, the parameter $n \alpha \approx 1/1300 $, and those corrections are negligible. Also, at detunings larger than the fine structure splitting of 10 GHz, optical pumping to other hyperfine states is suppressed. At 100 GHz detuning, the branching ratio is less than 1\% for any polarization of light, so no special cycling transition is needed. 
We used rather weak and long laser pulses with a Rayleigh scattering rate well below 1 photon/atom/ms to stay far away from nonlinear collective light scattering (see Methods). 

\begin{figure}
\includegraphics[width=\columnwidth]{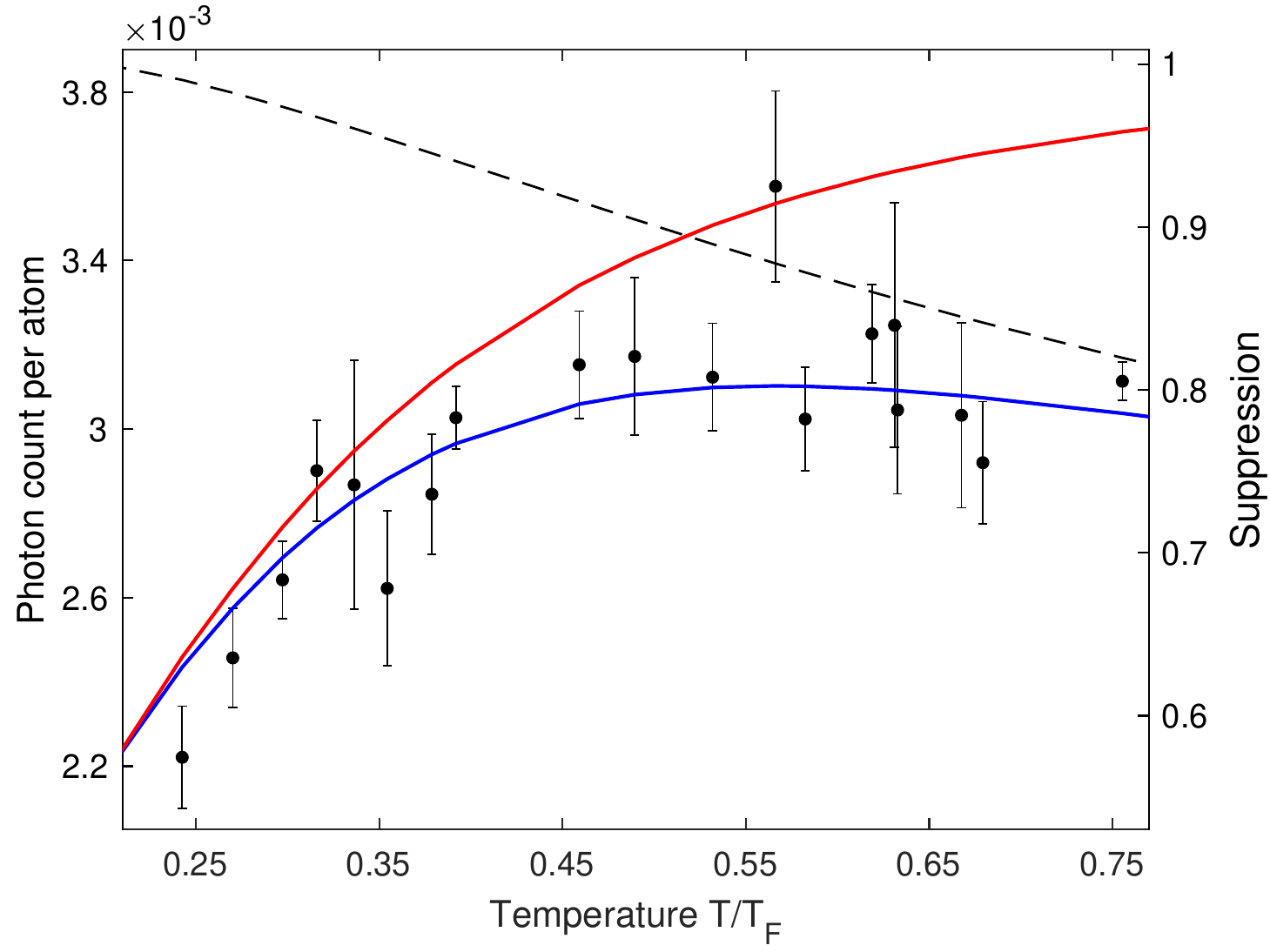}
\caption{\label{fig2} \textbf{Pauli blocking of light scattering.} Photon count per atom as a function of the cloud's temperature, observed at a scattering angle of 90 degrees. At low temperature, the scattered light is reduced by 40\% with respect to the unblocked case. The cloud is heated by turning the optical dipole trap off and on for a variable duration. The probe light is pulsed on for 25\,ms at a power of 2.35 mW and an intensity of $1.2\times10^4$\,mW/cm$^2$ and is detuned 100 GHz below the atomic resonance (at 671\,nm). The red curve is the theoretical prediction (Eq. 12 in \cite{Shuve_2009}) of the Pauli suppression factor (right y-axis).  The dashed black line shows the reduction of scattering due to the substantial thermal expansion of the cloud within the finite beam waist of the probe light (representing the integrated product of two Gaussians with rms widths of the cloud and the laser beam). The blue curve is the product of the red and black curves. There are no free fitting parameters apart from the overall vertical scale relative to the data. The error bars throughout the whole paper are purely statistical and reflect one standard error of the mean. Data points are averaged typically over 3 to 5 samples.}
\end{figure}

The preparation of our sample of approximately $6\times10^5$ lithium-6 atoms at temperatures $T/T_F \approx 0.2$ held in a far-off resonant dipole trap is summarized in \cite{ccaugri2020spin} and the Methods section. After cooling to the lowest temperature, the sample is heated either by strongly modulating the trapping potential or by scattering of light. For the observation of suppressed light scattering, we typically scatter 0.9 photons/atom during 25 ms, and collect the fluorescence at a right angle with an imaging system with a collection efficiency of 0.31\%. Figure \ref{fig2} shows the main result of this paper --- the suppression of light scattering by a degenerate Fermi gas. Our observations are in good agreement with theoretical calculations for a trapped cloud of atoms \cite{Shuve_2009} using the independently measured atom number and temperature and no adjustable fitting parameter besides the overall scaling. Results are limited to $T/T_F\le0.75$ in order to reduce systematic errors such as atom loss by spilling due to the finite trap depth and expansion of the cloud with respect to the probe beam.

\begin{figure}
\includegraphics[width=\columnwidth]{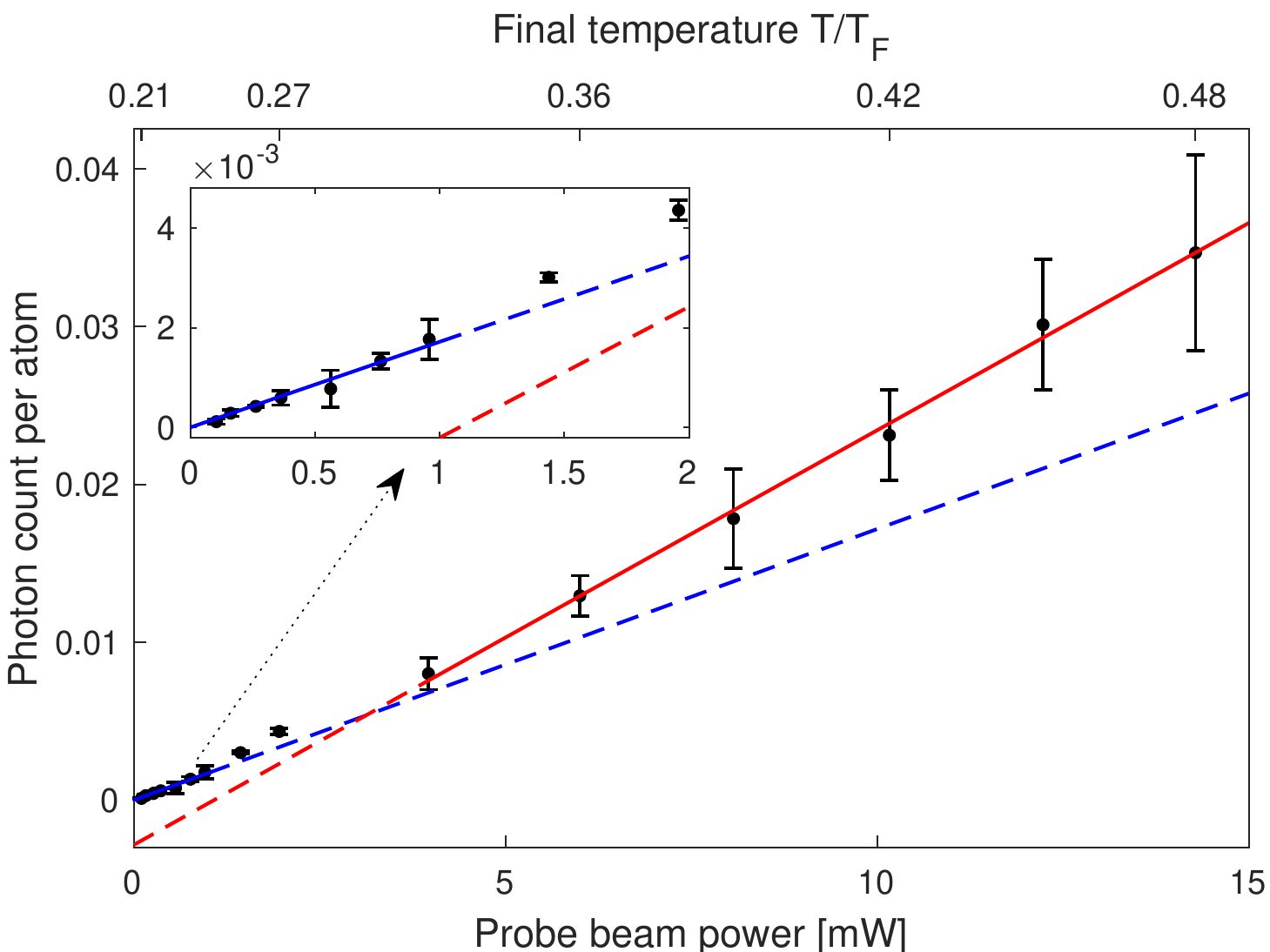}
\caption{\label{fig3} \textbf{Light scattering as a function of probe beam power.} We observe different slopes for low power (dashed blue line, with Pauli blocking) and high power (red line), where Pauli blocking is eliminated because of recoil heating of the cloud. The negative y-intercept of the red line represents the number of photons scattered at the Pauli suppressed rate in the early part of the 50 ms probe pulse. Top x axis shows the final temperature ($T/T_F$), measured after releasing atoms from the trap. Detuning of the probe beam is $\Delta = -112\,$GHz. Inset shows an enlarged version of the low power region. 
}
\end{figure}

Since light scattering heats up the cloud by photon recoil heating, Pauli suppression can be observed only for sufficient short or weak laser pulses. This is demonstrated in Fig. \ref{fig3} where we study the number of photons scattered per atom as a function of laser intensity. Pauli blocking shows up as the difference in the slope at low and at high intensities. Interestingly, the linear fit for high intensities does not go through the origin and intercepts the y-axis at a negative value. This value is proportional to the number of photons missing due to Pauli blocking during the initial part of the laser pulse.

Since light scattering involves a (virtual) excited state, fermionic suppression of light scattering is related to Pauli suppression of spontaneous emission from an excited state embedded in a Fermi sea. The distinction between light scattering and spontaneous emission becomes important for an interacting system. It was shown theoretically that spontaneous emission in a zero-temperature Bose-Einstein condensate is enhanced by bosonic stimulation via the quantum depletion, whereas light scattering from a Bose-Einstein condensate is suppressed since the static structure factor $S(q) < 1$ due to the phonon dispersion relation \cite{PhysRevA.63.041601}. 

So far, we have described Pauli blocking as a single particle effect due to Fermi statistics. However, since Fermi statistics creates correlations between particles, one can also express Pauli blocking in terms of a pair correlation function. This will allow us to compare the suppression in our light scattering experiment to other studies demonstrating fermionic suppression.
The cross section of light (and also particle) scattering with momentum transfer $\hbar q$, $d \sigma/d \Omega$ is given by $S(q)$ times the single particle cross section $\sigma_0(q)$:
$d \sigma/d \Omega = N \sigma_0(q) S(q)$ with $N$ the number of fermions. $S(q)$ is given by the Fourier transform of the density-density correlation function. A homogeneous system with $S(q)=0$ would not scatter light. Uncorrelated classical particles show Poissonian fluctuations implying $S(q) = 1$. Suppression of light scattering off fermions is caused by suppressed density fluctuations, implying $S(q) < 1$. Suppression of density fluctuations in cold fermion clouds has been directly observed in \cite{PhysRevLett.105.040402,PhysRevLett.105.040401}, where the atomic density was shown to have sub-Poissonian fluctuations. This immediately implies reduced light scattering at small angles of order $k_F/k_i$. We have now extended this work by suppressing light scattering at all angles, and directly detecting the scattered photons.
In the absence of longer-range correlations, the density-density correlation function is expressed by the pair correlation function $g(r)$ which is the normalized probability of detecting two particles at separation $r$, so the structure factor can be written as:
\begin{equation}
S(q) = 1 +n\int d^3\vec{r} ( g(r)-1) e^{-i\vec{q}\cdot \vec{r}}
\end{equation}

For a non-degenerate quantum gas, $g(r) -1 \ne 0$ for $r< \Lambda_t$ where $\Lambda_t$ is the thermal de Broglie wavelength \cite{PhysRevE.84.042101}:
\begin{equation}
g(r) \approx 1 \pm \exp(-2\pi r^2/\Lambda_t^2)
\end{equation}

resulting in 
\begin{equation}
\label{Eq3}
S(q)\approx 1\pm D\exp(-q^2\Lambda_t^2/8\pi)/2^{3/2}
\end{equation}
where $D=n\Lambda_t^3$ is the peak phase space density. The cloud averaged phase space density is $D/2^{3/2} $. The $\pm$ sign refers to bosons/fermions, respectively.
The term ``1’’ is the (normalized) contribution of the scattering of single atoms, whereas the second term is due to non-vanishing interference terms involving light scattering by pairs of particles. Or in other words, for a non-interacting Fermi (Bose) gases, the pair scattering interferes destructively (constructively) with individual particle scattering resulting in fermionic suppression (bosonic enhancement) which becomes strong only for high phase space densities near quantum degeneracy. Equation \ref{Eq3} can be generalized for degenerate gases \cite{PhysRevE.84.042101} with the result that for fermions at zero temperature $S(q\rightarrow0)=0$: pair scattering completely cancels the scattering from single particles at angles for which $q \ll k_F$. This description emphasizes the central role of the pair correlations in enhancement or suppression of light scattering. Similar effects happen for colloidal particles with spatial correlations due to interactions \cite{parola2014optical}. It is only for non-interacting gases that quantum degeneracy is necessary to strongly modify the structure factor and pair correlations are one-to-one related to bosonic enhancement or ferminonic suppression.

Fermionic anti-bunching ($g(r) < 1$) leads to other forms of Pauli suppression. It implies that at low temperatures collisions can only occur only by $p$-wave interactions, resulting in the suppression of elastic scattering \cite{PhysRevLett.82.4208} and inelastic scattering, including three-body recombination \cite{PhysRevLett.93.090404, ccaugri2020spin} which was crucial for the study of the BEC-BCS crossover, and the absence of interaction shifts in RF spectroscopy \cite{Gupta1723}. Sometimes, those processes are described by Pauli suppression in the input channel (which is $p$-wave), whereas the suppression of light scattering is regarded as Pauli suppression in the output channel. This distinction is correct, but obscures the common origin of both effects, which are the pair correlations.

\begin{figure}
\includegraphics[width=\columnwidth]{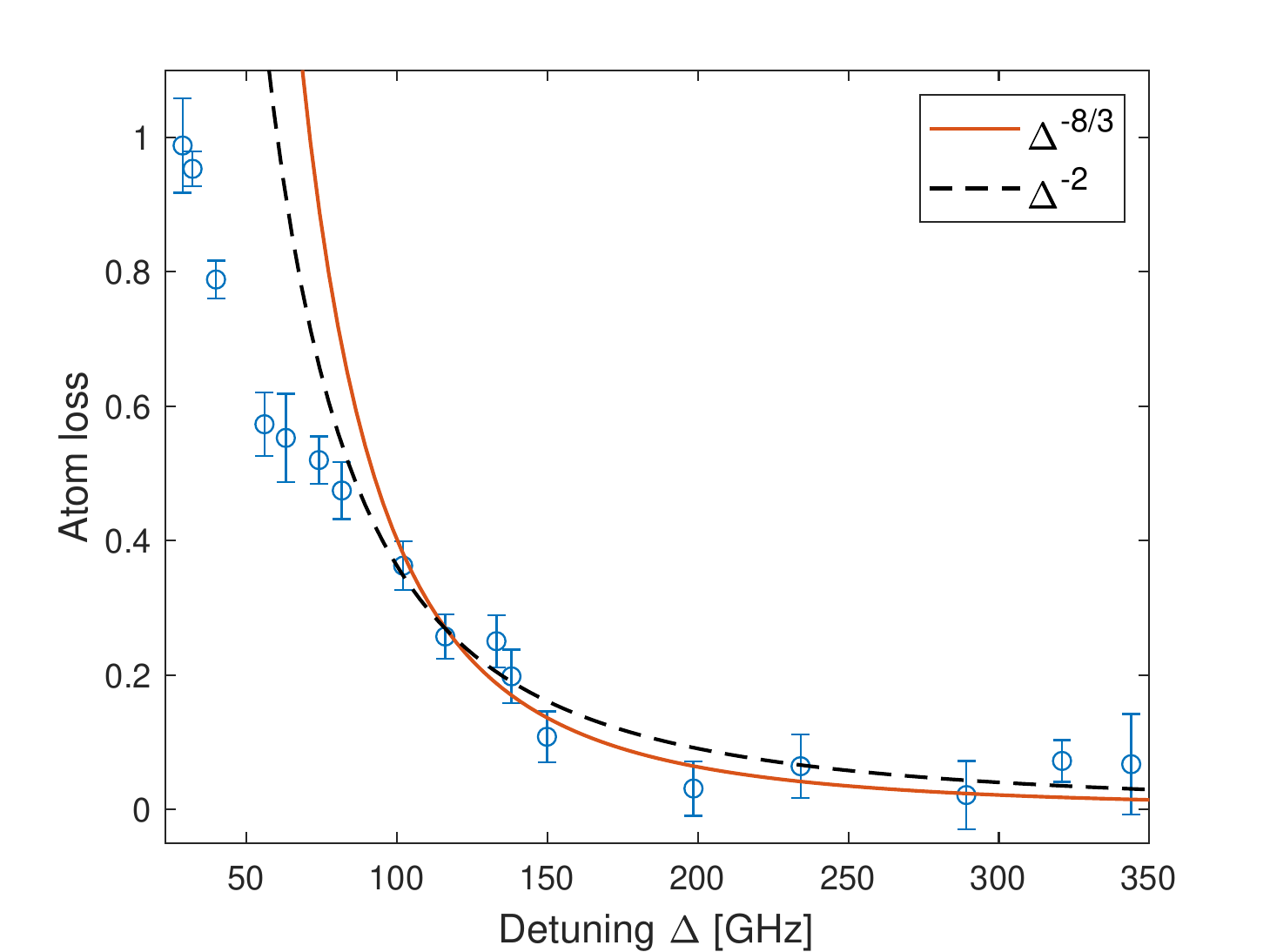}
\caption{\label{fig4} \textbf{Inelastic light scattering.} Shown is the atom loss versus blue detuning $\Delta$ of the probe laser. Atom loss is calculated as $-\log(N/N_0)$, where $N_0$ ($N$) is the number of atoms without (with) applying the probe beam. To guide the eye, we show power law dependence $\Delta^{-\alpha}$ for $\alpha=8/3$ (solid line) and 2 (dashed line), adjusted to the data points for $\Delta \ge$ 100 GHz. Exposure time is 25\,ms, and probe power is 2.1\,mW.}
\end{figure}

Suppression of elastic light scattering requires quantum degeneracy in order for the pair scattering to compete with the single particle scattering. In contrast, inelastic light scattering is suppressed at low temperatures independent of degeneracy, since it occurs only 
during collisions of two atoms.
We have studied Pauli suppression of inelastic light scattering with the goal of using it as a probe for the pair correlation function. The simplest limit for inelastic light scattering is far blue detuned light where pairs of colliding atoms are excited to a repulsive branch of the $1/r^3$ van der Waals potential. Resonant absorption takes place at a distance $r$ when $\Gamma/(r/\lambdabar)^3 \approx \Delta$. For detunings of tens of GHz, we are probing the pair correlations at distances $r \approx$ 10 nm, much smaller than the de Broglie wavelength of fermions.
For large detuning, inelastic light scattering is proportional to the pair correlation function at the resonant distance (Condon point): $p(r) r^2 (dr/ d \Delta)$ with $r \propto \Delta^{-1/3}$. For distinguishable particles, the pair correlation function is constant (with a cutoff at the interparticle spacing $n^{-1/3}$) resulting in an loss rate due to inelastic light scattering proportional to $\Delta^{-2}$. This result is well known and emerges from the Gallagher-Pritchard model of inelastic light scattering \cite{PhysRevLett.63.957} and was experimentally observed for rubidium-85 \cite{PhysRevA.54.R1030}. Fermions which have not been studied before, have a short range pair correlation function proportional to $r^2$ ($p$-wave scaling) and therefore a loss rate which scales as $\Delta^{-8/3}$. Our initial goal was to confirm this scaling. In Fig. \ref{fig4} we show the loss rate for ultracold fermions due to inelastic light scattering as a function of the (blue) laser detuning. Unfortunately, the decay did not follow a power law dependence for detunings smaller than 100 GHz. This is possibly related to to the lithium fine structure splitting of 10.05 GHz. Rubidium-87 didn't show a clean power law behavior, either, even beyond 100 GHz detuning \cite{PhysRevA.54.R1030}, possibly due to the large hyperfine splitting. We have not measured absolute loss rate coefficients, but estimate that bosonic atoms undergoing s-wave collisions would have a loss rate more than two orders of magnitude higher. In future work, we will either need molecular theory for smaller detunings or higher sensitivity at larger detuning to map out the fermionic pair correlation function. Red-detuned light has discrete photo-association resonances and small loss outside these resonances. This is why we use red detuning for the light scattering experiment.

In conclusion, we have directly observed Pauli blocking of light scattering. For our high densities, Pauli blocking is mainly limited by temperature, which can be lowered by an improved evaporation strategy addressing $p$-wave three body recombination as the dominant loss mechanism \cite{ccaugri2020spin}. Pauli suppression can be used in quantum simulations to create fermionic samples which are less sensitive to heating when probed or manipulated by laser light. There are still many unresolved questions in how dense atomic samples scatter light, involving dipole-dipole interactions and superradiant scattering \cite{PhysRevA.94.023612}, and fermionic clouds with reduced incoherent scattering are a promising system for further studies.

After this work was completed, we learned about two related studies of light scattering off fermions \cite{sanner2021pauli,deb2021observation}.

\bibliography{bibliography}


\section{Methods}

\textbf{Sample Preparation} Ultracold lithium clouds are prepared as in our previous work\,\cite{ccaugri2020spin}. In short, $^{23}$Na and $^{6}$Li atoms are first laser cooled, and transferred into a quadrupole magnetic trap with an optical plug \cite{PhysRevLett.75.3969}. Forced microwave evaporation of the Na atoms\,\cite{Zoran} sympathetically cools the lithium atoms. The lithium atoms are transferred into a single-beam 1064\,nm optical dipole trap (ODT) with variable spot size and power, which controls the trap volume and densities. A partially non-adiabatic RF Landau-Zener sweep transfers the majority of the atoms to the collisionally stable lowest Zeeman state $\ket{1/2,1/2} \equiv \ket{1}$, while keeping around 7\% in the original state $\ket{3/2,3/2} \equiv \ket{6}$. This creates a spin mixture with s-wave interactions, allowing for efficient evaporative cooling into quantum degeneracy. Decreasing the spot size of the trapping beam creates a tighter trap with frequencies of $\omega_r/2\pi = 34\,$kHz, $\omega_z/2\pi = 770\,$Hz. The atoms are exposed to a final stage of evaporative cooling by tilting the trapping potential with a magnetic gradient for 1.5\,sec. A typical sample contains $N\simeq 6\times10^5$ lithium-6 atoms at $T/T_F\simeq0.2$, with a Fermi temperature of $T_F=\hbar(\omega_r^2\omega_z 6N)^{1/3}=70\,\mu$K. This corresponds to a density of $\sim 6\times10^{14}$/cm$^3$ and an on-resonance optical density of $\sim$25,000.

By ramping the optical trap to full power we have generate degenerate samples with densities even up to $3\times 10^{15} \,\textrm{cm}^{-3}$ and Fermi energies of $190\,\mu$K. However, at these densities we observe fast three-body losses and associated heating (which occur even in a spin-polarized sample \cite{ccaugri2020spin}). At the final step before the experiment, the majority of the atoms are transferred by the same RF Landau-Zener sweep back to state $\ket{6}$, leaving $\leq$ 10 \% of the total number in state $\ket{1}$ to ensure thermalization. This transfer was implemented in the earlier phase of the experiment when we used smaller detunings because state $\ket{6}$ has a cycling transition. The number of atoms in the trap is measured using standard time-of-flight absorption imaging. We estimate the error in atom count to be no greater than 20\%.

\textbf{Light scattering and collection}
Red and blue far-detuned probe light is generated using a tunable titanium-sapphire laser. In the atom's plane, the $1/e^2$ diameter of the probe beam is 220$\,\mu$m. Scattered light is collected at a right angle with respect to the probe beam using a corrected lens (air-spaced triplet, NA=0.27) and detected with a low background noise CMOS camera (typical 1.3$e^-$ readout noise, quantum efficiency of 55\%). 
The total photon detection efficiency of 0.31\% includes the losses due to multiple filters required to block the collinear 1064\,nm trapping light. It was determined from the number of camera counts compared to the scattering rate calculated from laser power and detuning.

\begin{figure}
\includegraphics[width=\columnwidth]{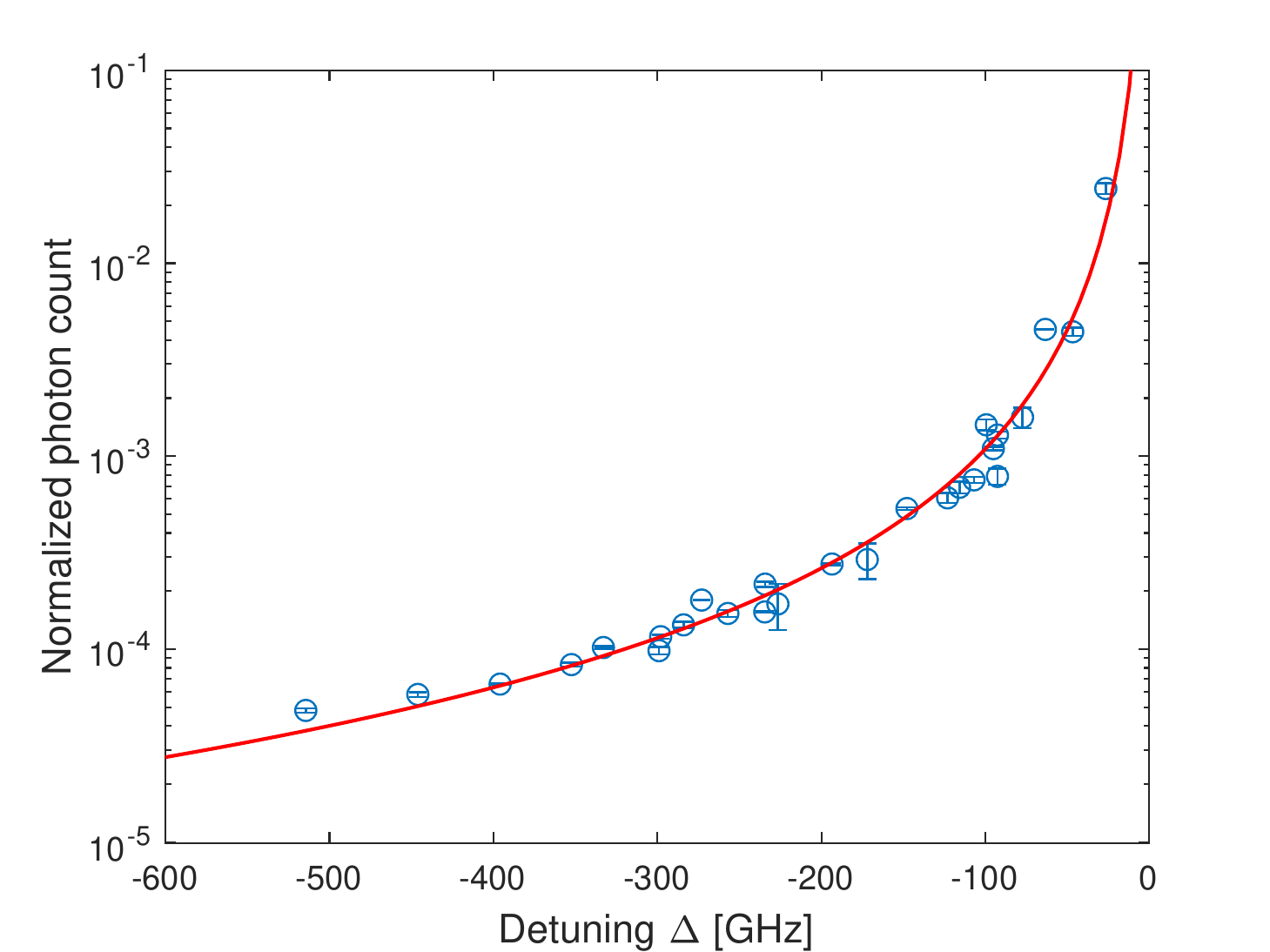}
\caption{\label{fig:DeltaSquare} \textbf{Light scattering cross section versus detuning $\Delta$.} Number of photons counts per atom as a function of red detuning of the probe beam, normalized to the beam power. A power law fit $\propto |\Delta|^\alpha$ returns a value of $\alpha=2.05 \pm 0.12$ (one $\sigma$ fit error). The excellent agreement with the off resonant $1/\Delta^2$ scaling confirms that at large detunings scattering is not sensitive to line shifts and broadening by collective light scattering. Detunings are with respect to the D2 line.}
\end{figure}

\textbf{Detuning of probe light}
Large laser detunings are necessary to avoid absorption, losses and dipolar corrections to the light scattering. However, due to the need for larger laser intensities at large detunings, stray light can become a problem. We therefore explored a large range of detunings. Blue detunings (see Fig. \ref{fig4}) cause much stronger losses than red detuning. For red detunings $\Delta$ of -20GHz to -500GHz, we have measured light scattering cross sections. The laser power was adjusted for almost constant scattering rate. Figure \ref{fig:DeltaSquare} shows that the measured cross sections have an exact $1/\Delta^2$ dependence indicating that we are safely in the single-particle light scattering regime, unaffected by possible line broadening due to collective light scattering, which in simple models is as large as the resonant optical density divided by 4 (in units of $\Gamma$) \cite{PhysRevA.94.023612}.
The Pauli blocking measurements were performed at $\Delta$ around -100 GHz. 

\textbf{Intensity of probe light}
Due to the high optical density (OD) of our sample, collective light scattering occurs at very low laser intensities, which is related to the gain process in recoil-induced resonances \cite{PhysRevLett.72.3017} or Rayleigh superradiance \cite{inouye1999superradiant}. This gain depends only on the Rayleigh scattering rate, independent of detuning. 
We checked that the photon scattering rate is linear to better than 10 \% for Rayleigh scattering rates up to 0.95 photon/atom/ms. The highest Rayleigh scattering rates for the data in Figs. \ref{fig2} and \ref{fig3} were smaller by factors of 25 and 4.3, respectively.

\textbf{Heating degenerate fermions}
After evaporative cooling to the lowest temperature $T/T_F \approx 0.2$, light scattering at higher temperatures was performed after heating up the sample by two different methods. (1) The optical dipole trap was switched off for a variable duration and then back on, either once or multiple times. (2) Photon recoil heating by an extra light pulse for a variable duration before probing for light scattering. The heating was followed by a hold time of 250\,ms for thermalization. Either of the two methods was used to obtain data on Pauli suppression as shown in Fig. \ref{fig2}. Due to slow p-wave collisions, heated clouds were slightly out of equilibrium as evidenced by a small anisotropy in time-of-flight expansion.

\vbox{}
\noindent \textbf{\large Acknowledgments} \\
We thank Hyungmok Son for comments on the manuscript. We acknowledge support from the NSF through the Center for Ultracold Atoms and Grant No. 1506369, ARO- MURI Non-Equilibrium Many-Body Dynamics (Grant No. W911NF-14-1-0003), AFOSR- MURI Quantum Phases of Matter (Grant No. FA9550- 14-10035), ONR (Grant No. N00014-17-1-2253), and from a Vannevar-Bush Faculty Fellowship.

\vbox{}
\noindent \textbf{\large Author Contributions} \\
All authors contributed to the concepts of the experiment, Y.M. and F.T. developed the sample preparation, Y.M and Y.L. designed the light scattering setup, took and analyzed the data, Y.M, Y.L. and W.K. wrote the paper.

\vbox{}
\noindent \textbf{\large Competing interests} \\ 
The authors declare no competing interests.


\noindent \textbf{Correspondence and requests for materials} should be addressed to margalya@mit.edu.

\vbox{}
\noindent \textbf{\large Data Availability} \\ The data that support the plots within this paper and other findings of this study are available from the corresponding authors upon reasonable request.


\end{document}